\documentclass[%
 reprint,
 amsmath,amssymb,
 aps,
prb,
superscriptaddress,
floatfix,
10pt]{revtex4-2}

\usepackage{graphicx}
\usepackage{dcolumn}
\usepackage{hyperref}
\hypersetup{
    colorlinks=true,
    linkcolor=blue,
    filecolor=magenta,      
    urlcolor=blue,
    citecolor=blue,
    allcolors = blue
    }
\usepackage{physics}
\usepackage[mathlines]{lineno}
\usepackage{amsmath}
\makeatletter
\g@addto@macro\normalsize{%
  \setlength\abovedisplayskip{12pt}
  \setlength\belowdisplayskip{12pt}
  \setlength\abovedisplayshortskip{12pt}
  \setlength\belowdisplayshortskip{12pt}
}
\makeatother
\usepackage{bm} 
\usepackage{amssymb}
\usepackage{multirow}
\usepackage{tabularx}
\usepackage{array}
\usepackage[version=4]{mhchem}

\usepackage{soul}

\begin{document}


\title{Many-body effects on the quasiparticle
band structure and optical response of single-layer penta-\ce{NiN2}}
\author{Enesio Marinho Jr.}
 \email{enesio.marinho@unesp.br}
\affiliation{Departamento de Física e Química, Universidade Estadual Paulista (UNESP),\\ Av.\ Brasil, 56, Ilha Solteira, 15385-007 São Paulo , Brazil}
\affiliation{
 Instituto de Física Teórica, Universidade Estadual Paulista (UNESP),\\
R.~Dr.~Bento Teobaldo Ferraz, 271, 01140-070 São Paulo, São Paulo, Brazil.}
\author{Cesar E. P. Villegas}
\email{cesarperezvillegas@gmail.com}
\affiliation{
 Departamento de Ciencias, Universidad Privada del Norte, Lima 15434, Peru}
 \affiliation{
  Instituto de Física Teórica, Universidade Estadual Paulista (UNESP),\\
R.~Dr.~Bento Teobaldo Ferraz, 271, 01140-070 São Paulo, São Paulo, Brazil.}
\author{Pedro Venezuela}
\email{pedrovenezuela@id.uff.br}
\affiliation{Instituto de Física, Universidade Federal Fluminense (UFF),\\ Av.~Gal.~Milton Tavares de Souza, s/n, 24210-346 Niterói, Rio de Janeiro, Brazil.}
\affiliation{
 Instituto de Física Teórica, Universidade Estadual Paulista (UNESP),\\
R.~Dr.~Bento Teobaldo Ferraz, 271, 01140-070 São Paulo, São Paulo, Brazil.}
\author{Alexandre R. Rocha}
\email{alexandre.reily@unesp.br}
\affiliation{
  Instituto de Física Teórica, Universidade Estadual Paulista (UNESP),\\
R.~Dr.~Bento Teobaldo Ferraz, 271, 01140-070 São Paulo, São Paulo, Brazil.}

\date{\today}

\begin{abstract}
We present a comprehensive first-principles study on the optoelectronic properties of the single-layer nickel diazenide (penta-\ce{NiN2}), a recently synthesized Cairo pentagonal 2D semiconductor. We carry out \textit{ab initio} calculations based on the density-functional theory (DFT) and many-body perturbation theory, within the framework of Green's functions, to describe the quasiparticle properties and analyze the excitonic effects on the optical properties of monolayer penta-\ce{NiN2}. Our results reveal a quasiparticle band gap of approximately 1 eV within the eigenvalue self-consistent $GW$ approach, corroborating the monolayer penta-\ce{NiN2}'s potential in optoelectronics. Remarkably, the acoustic phonon-limited carrier mobility for the monolayer penta-NiN2 exhibits an ultra-high hole mobility of $84{\times}10^4$ cm$^2$/V$\cdot$s. Furthermore, our findings indicate that the material's band gap exhibits an anomalous negative dependence on temperature. Despite being a two-dimensional material, monolayer penta-{NiN2} presents resonant excitons in its most prominent absorption peak. Therefore, penta-\ce{NiN2} boasts compelling and promising properties that merit exploration in optoelectronics and high-speed devices.
\end{abstract}

\maketitle


\section{\label{sec:introduction}Introduction}
Motivated by the prediction of penta-graphene \cite{zhang2015PG}, a 2D carbon allotrope composed entirely of pentagons, much effort has been devoted to discovering novel pentagon-based 2D materials with intriguing properties. Since then, more than a hundred pentagon-based 2D systems have been proposed \cite{Shen2022REVIEW}, such as the unitary sheets of penta-silicene \cite{pentasilicene1,penta-silicene-PRA}, penta-germanene \cite{penta-germanene}, and penta-tellurene \cite{penta-tellurene}, as well as the binary sheets penta-\ce{MS2} (M = Ni, Pd, Pt) \cite{penta-XS2}, penta-\ce{PdSe2} \cite{penta-PdSe2}, among others \cite{penta-CN2, penta-B2C,li2016penta-BN2, SiX2,penta-BCN,penta-BCP,penta-CNP,PdPX, PtPX,FeAsS,PdSeX,PdSeX,Zn2C2P2}.

Some of those pentagonal-based materials have already been synthesized. Pentagonal silicene nanoribbons (NRs), for example, were synthesized by Cerdá \emph{et al.} \cite{Cerda2016pentaSinanoribbon}, who grew these 1D structures on a Ag(110) surface. Nevertheless, the first pentagon-based 2D sheet successfully synthesized was the penta-PdSe2, which was obtained by exfoliating its bulk counterpart \cite{OyedelePdSe2}. Combining physical and chemical vapor deposition techniques, Zhang \emph{et al.} \cite{zhang2021PdS2} reported the experimental realization of few-layered penta-\ce{PdS2}. Regarding ternary systems, the penta-PdPSe single-layer was successfully fabricated by high-temperature solid-state reaction \cite{li2021PdPSe}. Also, Wang \emph{et al.} \cite{wang2022PdPS}  have synthesized penta-PdPS monolayer and few-layer via chemical vapor transport methods.

\begin{figure}[t!]
    \centering
    \includegraphics[width=.92\linewidth]{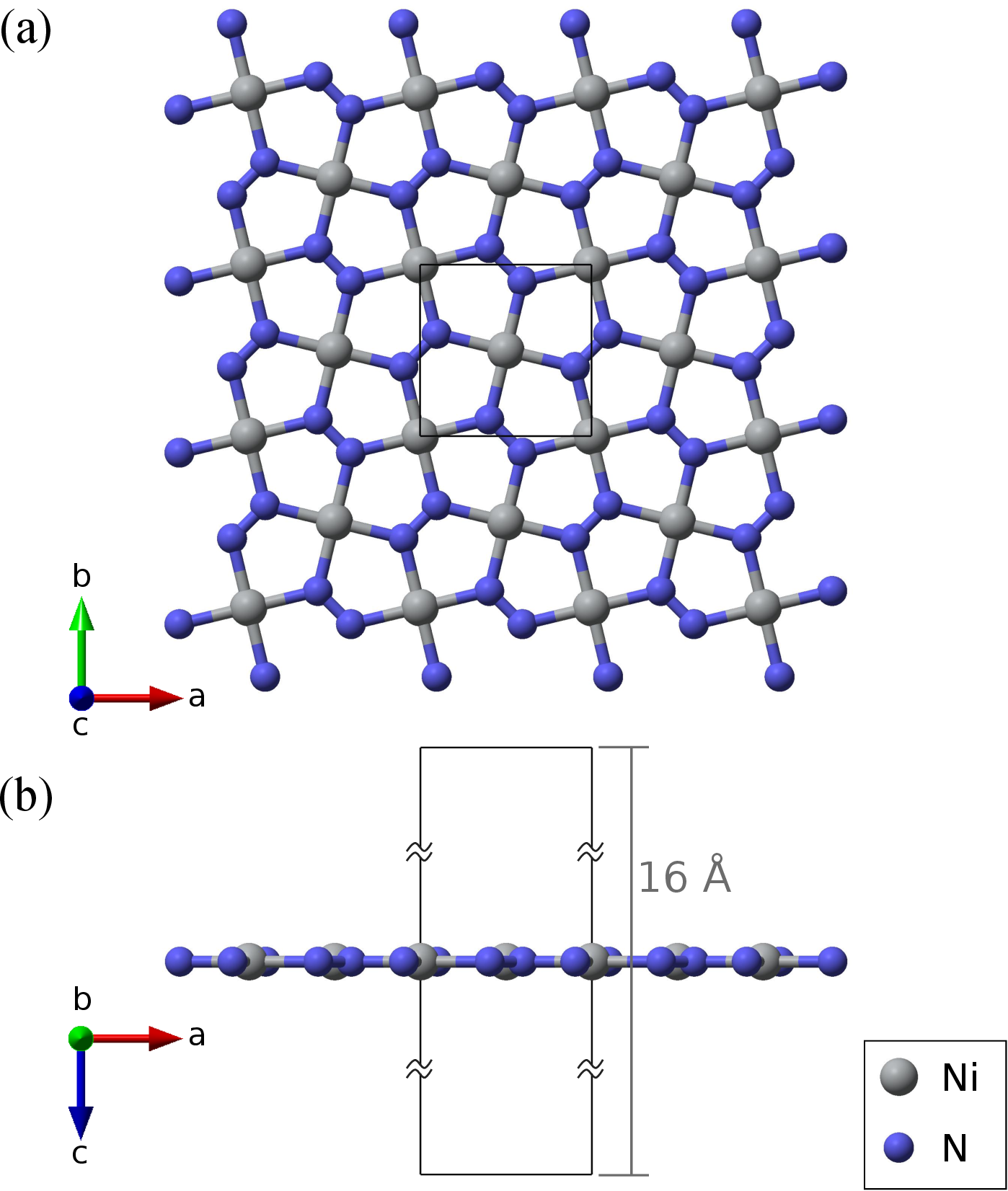}
    \caption{(Color online) (a) Top and (b) side views of the crystal structure of monolayer penta-{NiN2}, with Ni and N atoms being represented by grey and blue spheres, respectively. The square unit cell is represented by black solid lines.}
    \label{fig:structure}
\end{figure}

Employing a combination of crystal-chemical design and high-pressure bottom-up approach, Bykov \emph{et al.} \cite{Bykov2021} have synthesized pentagonal nickel diazenide (penta-\ce{NiN2}), whose crystal structure is composed of distorted pentagonal motifs forming an ideal prototype of Cairo tessellation (Fig.~\ref{fig:structure}). Theoretical calculations have predicted that this novel 2D pentagonal semiconductor exhibits remarkably high elastic modulus, tensile strength, and room-temperature lattice thermal conductivity \cite{mortazavi2022outstanding}. Moreover, first-principles calculations indicated that the single-layer penta-\ce{NiN2} is a direct band-gap semiconductor, while the bulk phase is metallic \cite{Bykov2021}. Thus, at the PBE-DFT level of theory, its electronic band gap yielded about $0.05$ eV \cite{ApSurfSci-YUAN,mortazavi2022outstanding}, whereas at the hybrid functional HSE06 level, its band gap increases to $1.1$ eV, suggesting that the monolayer penta-{NiN2} holds potential for optoelectronic applications \cite{ApSurfSci-YUAN,mortazavi2022outstanding, Bykov2021}. Despite this, experimental measurements of the fundamental band gap for single-layer penta-\ce{NiN2} are still lacking. Hence, accurate predictions of its electronic and optical properties based on many-body perturbation theory (MBPT) can provide useful insights. 

Herewith, we analyze the role of the electron-electron and electron-hole interactions in the electronic band gap renormalization and the optical properties of penta-\ce{NiN2}. Our quasiparticle band structure within eigenvalue self-consistent $GW$ indicates that the monolayer penta-{NiN2} is a direct band gap semiconductor, with a band gap magnitude of $1.05$ eV. Moreover, our results indicate that this pentagonal 2D semiconductor presents ultrahigh hole mobility, and a negative pressure coefficient which suggests an anomalous temperature dependence of the band gap. Finally, optical absorption has shown that excitonic effects play an important role in its spectrum, with the presence of dark excitons and bright resonant states. 

\begin{figure*}[t!]
    \centering
    \includegraphics[width=\linewidth]{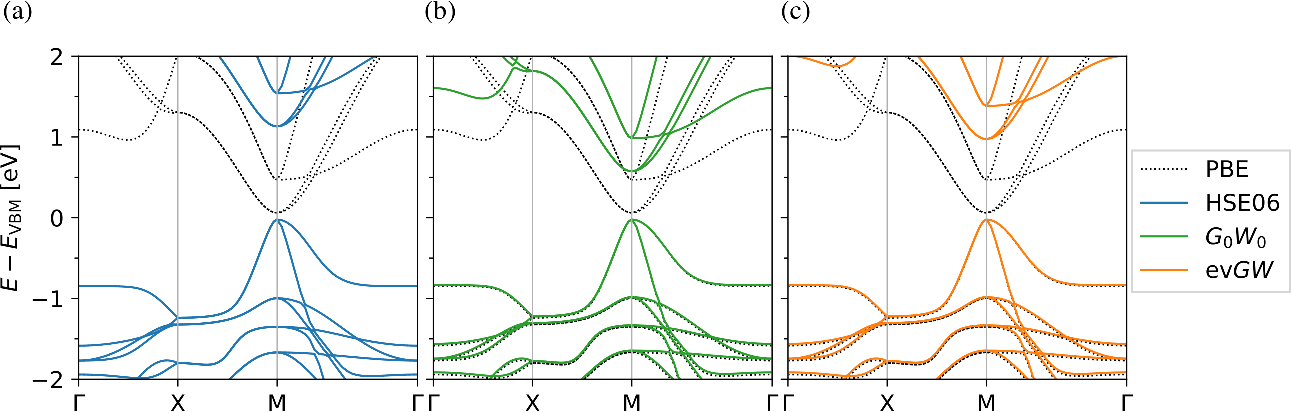}
    \caption{(Color online) Electronic band structure of monolayer penta-{NiN2} computed within (a) HSE06, (b) $G_0W_0$ and (c) ev$GW$. The DFT-PBE results (dashed black lines) are shown for comparison. The energies refer to the valence band maximum.}
    \label{fig:gw-bands_pentaNiN2}
\end{figure*}
\section{\label{sec:method}Computational Methods}
The ground-state structural and electronic properties have been obtained by
means of density-functional theory (DFT) \cite{HohenbergKohn1964,KohnSham1965}, as implemented in the \textsc{q}uantum \textsc{espresso} package \cite{qe}. The Perdew-Burke-Ernzerhof generalized-gradient approximation (GGA-PBE) \cite{pbe} and the screened hybrid functional of Heyd, Scuseria, and Ernzerhof (HSE06) \cite{hse06-1st,hse06-2nd} were employed to describe the exchange-correlation functional. Fully relativistic ONCVPSP \cite{ONCVPSP} norm-conserving pseudopotentials are used to describe the electron–ion interaction. Equilibrium structural properties have been computed by fully relaxing both the in-plane unit cell and the atomic positions, with a force convergence criterion of $1{\times}10^{-5}$ Ry/Bohr. A vacuum spacing of ${\sim}16$ \AA{} was used in the out-of-plane direction to avoid spurious interaction with the periodic image of the system. A kinetic energy cutoff of 96 Ry was adopted to expand the Kohn-Sham orbitals in a plane-wave basis set. The charge density has been calculated by sampling the Brillouin zone with a Monkhorst-Pack $\Gamma$-centered \emph{k}-mesh of $14 {\times} 14 {\times} 1$. 

Quasiparticle energies are calculated as corrections to the Kohn–Sham eigenvalues using the Green’s function formulation of MBPT. Taking the first order Taylor expansion of the self-energy $\Sigma_{n\vb{k}}$ around the Kohn-Sham eigenvalues $\epsilon_{n\mathbf{k}}^{\text{KS}}$ (Newton
approximation), we have 
\begin{equation}
E_{n\mathbf{k}}^{\text{QP}} = \epsilon_{n\mathbf{k}}^{\text{KS}} +Z_{n\vb{k}} \mel{n\mathbf{k}^{\text{KS}}}{\Sigma_{n\vb{k}}(\epsilon_{n\vb{k}}^{\text{KS}}) -V^{\text{xc}}_{n\mathbf{k}}}{n\mathbf{k}^{\text{KS}}}\,,
\end{equation}
where $Z_{n\vb{k}} = \left[1 - \eval{\dv*{\Sigma_{n\vb{k}}(\omega)}{\omega}}_{\omega=\epsilon_{n\vb{k}}^{\text{KS}}} \right]^{-1}$ is the quasiparticle renormalization factor, $\ket{n\mathbf{k}^{\text{KS}}}$ describes the Kohn-Sham eigenstates, $V^{\text{xc}}_{n\mathbf{k}}$ is the DFT exchange-correlation potential. 
In the $G_0W_0$ approximation \cite{hedinGW}, the self-energy is described by $\Sigma = iG_0W_0$, corresponding to the first iteration of Hedin’s equations \cite{golze2019FC}. The \textsc{yambo} code \cite{yambo1,yambo2} was used to perform these calculations. 
We employ an energy cutoff of 100 Ry for the exchange part of the self-energy. For the correlation part of the self-energy,
the dielectric screening matrix is constructed considering the Plasmon-Pole approximation \cite{Godby-Needs-PhysRevLett.62.1169} with a $\vb{G}$-vector energy cutoff of 24 Ry and including 800 bands. 
The GW sum-over-states is performed on 1800 bands. 
This number of bands is sufficient to provide accurate results owing to the implementation of the Bruneval-Gonze terminators approach \cite{BGterminatorPRB2008}. 

To reduce spurious interactions among periodic images in the out-of-plane direction, we described the Coulomb potential as a truncated potential in a slab geometry with a cutoff of 8 Ry. Furthermore, to speed up the convergence of the $GW$ calculations w.r.t the $k$-point sampling, we employ the W-av method \cite{guandalini2023-wav-method}, which combines Monte Carlo integration techniques with an interpolation scheme of the screened potential. We have set the $\vb{G}$-vectors cutoff to 1.2 Ry. According to our convergence tests (see Supplementary Information), adopting the described parameters one can compute the quasiparticle band gap of the monolayer penta-{NiN2} within a convergence threshold of $0.1$ eV.

To overcome some issues associated with the rather pronounced starting-point dependence of the $G_0W_0$ approach, mainly regarding semi-local functional-based narrow band gap semiconductors, we also compute the quasiparticle energies within the eigenvalue self-consistent $GW$ (ev$GW$). In this approach, the input Kohn-Sham eigenvalues are replaced by the quasiparticle energies from the previous iteration, and the updated eigenvalues are used to construct $G$ and $W$, until the self-consistency in the quasiparticle energies is reached. Our convergence analysis for the ev$GW$ calculations indicated that self-consistency has been achieved after 3 iterations.  


The excitonic effects on optical spectra are analyzed by solving the Bethe-Salpeter equation (BSE) in the resonant (Tamm-Dancoff) approximation \cite{tamm1945,dancoff1950}, also employing the \textsc{yambo} code \cite{yambo1,yambo2}. The BSE Hamiltonian is given by
\begin{equation}
H^{\text{BSE}}_{\substack{vc\vb{k}\\v'c'\vb{k}'}} = \left(\epsilon^{\text{QP}}_{c\mathbf{k}}-\epsilon^{\text{QP}}_{v\mathbf{k}}\right)\var_{c,c'}\var_{v,v'}\var_{\vb{k},\vb{k}'}+ \Xi^{\text{eh}}_{vc\vb{k},v'c'\vb{k}'}\,,
\end{equation}
where the first term in parenthesis comprises a diagonal part that contains the
quasiparticle energy differences, and $\Xi^{\text{eh}}= K^{x} + K^{c}$ is the so-called BSE kernel, composed of an $e$-$h$ attraction term ($K^{c}$) and a repulsive exchange ($K^{x}$)\cite{louie2000PRB}.

Thus, by diagonalizing the BSE Hamiltonian,
\begin{equation}
\sum_{v'c'\vb{k}'}H^{\text{BSE}}_{\substack{vc\vb{k}\\v'c'\vb{k}'}}A^{S}_{v'c'\vb{k}'} = \Omega_SA^{S}_{vc\vb{k}}\,,
\end{equation}
we obtain the exciton envelope function $A^S_{vc\mathbf{k}}$ and its eigenvalues $\Omega_S$, for each exciton $S$.

Converged optical absorption spectra
have been obtained including the three highest-occupied
valence bands and the three lowest-unoccupied conduction bands, with a fine $\vb{k}$-grid sampling of $40 {\times}40{\times} 1$.

The optical response can be analyzed by the macroscopic dielectric function, which is computed using the exciton eigenstates and eigenenergies as follows
\begin{eqnarray}
    \epsilon_M(\omega) &=& 1-\lim_{\vb{q}\to 0}\frac{8\pi e^2}{\abs{\vb{q}}^2VN_q}\sum_{\substack{vc\vb{k}\\v'c'\vb{k}'}} \rho^*_{vc\vb{k}}(\vb{q},\vb{G})\rho_{v'c'\vb{k}'}(\vb{q},\vb{G}')\nonumber\\ &\times&\sum_{S} \frac{A^S_{vc\vb{k}}\left(A^S_{v'c'\vb{k}'}\right)^*}{\omega-\Omega_S}\label{eq:epsilon2}
\end{eqnarray}
where $\rho_{vc\vb{k}}(\vb{q},\vb{G}) = \mel{v\vb{k}}{e^{i(\vb{q}+\vb{G})\cdot\vb{r}}}{c\vb{k}-\vb{q}}$ are the single-particle dipole matrix elements, $\vb{G}$ is the reciprocal lattice vector, $N_q$ is number of transferred momenta $\vb{q}$, $V$ is the real-space unit cell volume.



However, it is noteworthy to stress that when employing 
the Coulomb truncation scheme, the real and imaginary part of the macroscopic dielectric function goes to unity and zero, respectively. This is a consequence of simulating an atom-thick structure that is fully surrounded by a vacuum. In this regard, the imaginary part of the polarizability per unit area ($\alpha_{2}$) is related to the optical absorbance \cite{pulciAPL2012,bernardi2013extraordinary}
\begin{equation}
    A(\omega) = \frac{4\pi\omega}{c}\alpha_2(\omega)\,.
\end{equation}

Hereafter, the optical absorption spectrum of the monolayer penta-{NiN2} will be described using the absorbance as the descriptive observable.

\section{\label{sec:results}Results and Discussion}

The relaxed crystal structure of penta-\ce{NiN2} is shown in Fig.~\ref{fig:structure}. It consists of an atomic-thick planar sheet composed of \ce{Ni2N3} pentagons, with tetragonal P$4/mbm$ symmetry (space group no. 127).   Our calculated lattice parameters are
$a=b=4.53$ \AA{}, and the bond lengths of
the \ce{Ni-N} and \ce{N-N} bonds are $1.875$ \AA{} and $1.246$ \AA{}, respectively. These values reproduce the experimental structural parameters \cite{Bykov2021} with errors below 1.5\%, and are also consistent with previous theoretical results \cite{ApSurfSci-YUAN,jpclett-pNiN2}.

The electronic band structure of penta-\ce{NiN2} was first computed at the DFT-PBE level. We obtained a direct bandgap of 0.07 eV, which agrees with prior theoretical results that indicate a PBE band gap smaller than 0.1 eV \cite{ApSurfSci-YUAN,jpclett-pNiN2}. We also calculated the electronic band gap using the hybrid HSE06 functional, obtaining a magnitude of 1.14 eV. This result is also in close agreement with previously reported values of $\sim 1.1$ eV \cite{ApSurfSci-YUAN,jpclett-pNiN2}. 

We further proceeded to compute the quasiparticle band structures of penta-\ce{NiN2} at both $G_0W_0$ and ev$GW$ levels, as shown in Fig.~\ref{fig:gw-bands_pentaNiN2}. The obtained $G_0W_0$ bandgap was 0.65 eV, whereas the quasiparticle band gap within ev$GW$ yielded 1.05 eV. 
The PBE-DFT band structure is also shown for comparison. 

In general, the MBPT formalism provides a suitable and accurate framework to incorporate the electronic correlation beyond the single-particle mean-field theories. However, the $G_0W_0$ results exhibit a dependence on the description of the starting Kohn-Sham eigensystem. To clarify this starting-point dependence, we also compare the monolayer penta-{NiN2} results to the reported results for germanium \cite{gantPRMater2022}.

The GGA-PBE band gap of the germanium is 0.08 eV, substantially underestimated and with the same magnitude that we obtained for the penta-\ce{NiN2} at this level of theory. In addition, the quasiparticle band gap of the germanium computed within $G_0W_0$ approach is still underestimated, yielding 0.47 eV. Thus, other $GW$ flavors should be employed. The eigenvalue self-consistent $GW$ (ev$GW$) is noticeably less dependent on the DFT starting point, providing a higher quasiparticle band gap of 0.95 eV , which is expected owing to the self-consistent update of the quasiparticle corrections.    

\begin{figure}[!t]
    \centering
    \includegraphics[width=.45\textwidth]{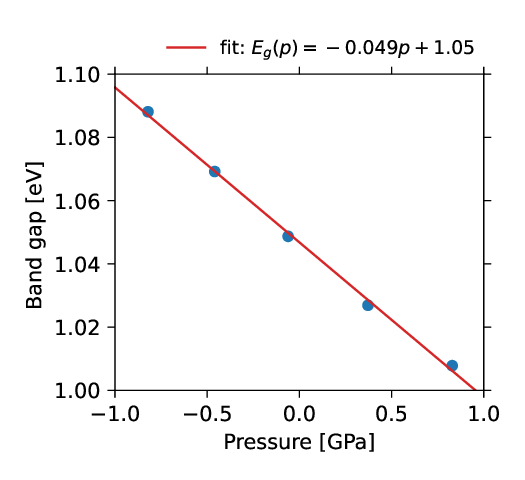}
        \caption{Electronic band gap as a function of the pressure for the monolayer penta-{NiN2}. We apply the quasiparticle correction within ev$GW$ approach as a scissor operator $\Delta_{\text{QP}}=0.97 $ eV to the PBE-DFT band gaps.}
        \label{fig:eg-pressure}
\end{figure}


As a result, we expect that the experimental band gap of the penta-\ce{NiN2} should be about 1 eV, close to the result yielded within the ev$GW$ approach, indicating that this pentagon-based 2D semiconductor is indeed a promising candidate for optoelectronics. 

\begin{figure*}[t!]
    \includegraphics[width=\textwidth]{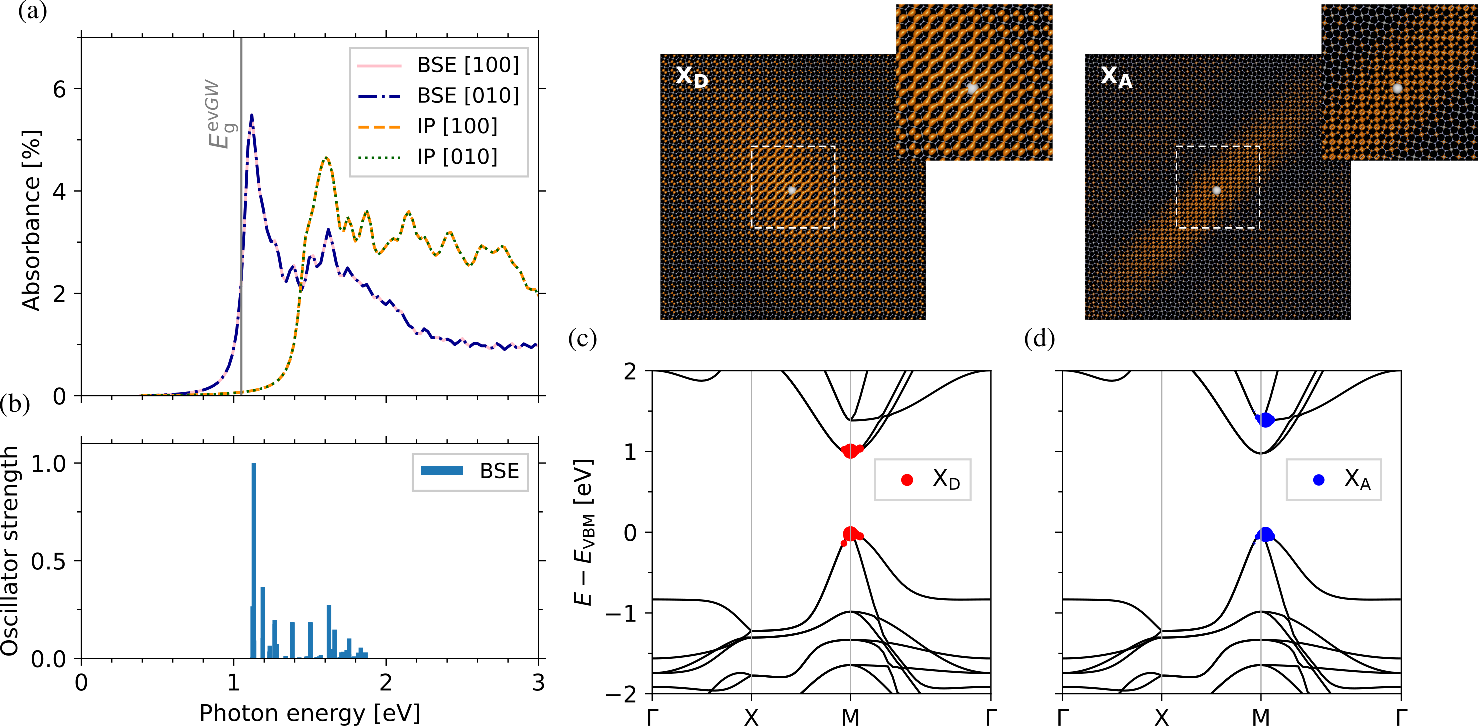}
   \caption{(Color online) Optical properties of the monolayer penta-{NiN2}. (a) Absorbance along [100] and [010] directions including excitonic effects by solving the Bethe-Salpeter equation (BSE) and at the independent particle (IP) level. (b) Oscillator strengths of the optical transitions normalized to 1. Exciton wavefunctions (top panels) and exciton weights on the Brillouin zone (bottom panels) for (c) the first dark exciton (X\textsubscript{D}) and (d) the most intense bright exciton (X\textsubscript{A}). The isosurface in orange ($0.005$ e/\AA$^3$) represents the electron density associated with the hole at a fixed N site (white spheres).}
    \label{fig:absorbance-oscillator_pNiN2}
\end{figure*}



We turn to analyze the carrier mobilities in single-layer penta-\ce{NiN2}. For 2D systems with isotropic effective mass \cite{langPRB2016}, the intrinsic mobility limited by acoustic-phonon scattering can be estimated through Bardeen–Shockley deformation potential theory \cite{bardeen-DPT} as follows 
\begin{equation}
    \mu_{2\text{D}} = \frac{e\hbar^3C_{\text{2D}}}{k_BT\abs{m_i^*}^2\left(\Xi_{1}\right)^2}\,,\label{eq:mobility}
\end{equation}
where $C_{\text{2D}}$ is the longitudinal elastic constant, extracted from the relation $(E - E_0)/S_0 = C_{2\text{D}}\delta^2
/2$, with $E$ being the total energy for the strained system, $E_0$ and $S_0$ the total energy and the surface area for the system at equilibrium, and $\delta=\Delta l/l_0$ is the applied uniaxial strain. $\Xi_{1}$ is the acoustic deformation potential defined as $\Xi_{1} = \Delta E_{\text{b}}/\delta$, where $\Delta E_{\text{b}}$ describes the energy shifts in the valence or conduction band edge, under proper compressive or tensile strain. We have computed those features applying a uniaxial strain $\delta$ varying from $-2\%$ to $2\%$, in steps of $0.5\%$, along the direction [110]. The calculated carrier effective masses, longitudinal elastic constant, and mobilities are summarized in Table \ref{tab:mobilities}. The obtained longitudinal elastic constant, $C_{\text{2D}}=175.9$ N/m, is in close agreement with other reported theoretical results of $172.2$ N/m \cite{zhang2022effect} and 174.9 N/m \cite{ApSurfSci-YUAN} .  

\begin{table}[t!]
    \caption{ Predicted carrier mobilities of monolayer penta-{NiN2} and the relevant parameters [see Eq.~\eqref{eq:mobility}] along the [110] direction. The results have been calculated at $T = 300$ K, within the DFT-PBE level. }
    \label{tab:mobilities}
    \begin{ruledtabular}
    \centering
    \begin{tabular}{lcccc}
      Carrier  & $m^*$ ($m_0$)& $C_{\text{2D}}$ (N/m) &$\Xi_{1}$ (eV) & $\mu$ $\left(\times 10^4 \text{ cm}^{2}/\text{V$\cdot$ s}\right)$  
 \\\hline\\[-.3cm]
      $e$ &0.34 & 175.9 &2.72 & 0.44 \\
      $h$ & 0.097 & 175.9 & 0.69 & 83.6 \\
    \end{tabular}
    \end{ruledtabular}
\end{table}

The computed mobilities yielded ultra-high hole mobility of ${\sim}84\times10^4$ cm$^2$/V$\cdot$s, which is one or two orders of magnitude higher than that of electrons, owing to the lowest effective mass and deformation potential for holes. These results are similar to the carriers mobilities of other pentagonal-based semiconductors, such as penta-\ce{PdP2}, with $\mu_h \sim 84\times 10^4$ cm$^2$/V$\cdot$s, and penta-\ce{PdAs2}, with $\mu_h \sim 89\times 10^4$ cm$^2$/V$\cdot$s \cite{yuan2019planar}. Again for those materials, the electron mobilities are two orders of magnitude smaller than the hole mobilities. The remarkably high carrier mobility of monolayer penta-{NiN2} holds great potential for high-speed electronic devices.

In addition, we analyzed the pressure dependence of the band gap for monolayer penta-{NiN2}. The effect on the gap due to the contraction of the lattice with decreasing temperature is intimately related to the response of the electronic band structure upon applying an external hydrostatic pressure \cite{francisco-lopezJPCL2019}. The thermal
expansion effects (TE) on the band gap can be estimated by
\begin{equation}
    \left(\pdv{E_g}{T}\right)_{\text{TE}} = -\alpha_VB_0\dv{E_g}{P}\,,\label{eq:pressure-coefficien}
\end{equation}
where $-\alpha_V$ is the volumetric expansion coefficient, $B_0$ is the bulk modulus, and $\dd E/\dd P$ is the so-called pressure coefficient of the gap \cite{francisco-lopezJPCL2019}. At room temperature, $\alpha_V$ is positive, and the sign of Eq.~\eqref{eq:pressure-coefficien} for the thermal expansion contribution depends on the sign of the
pressure coefficient. For most direct bandgap semiconductors, the pressure coefficient is positive, and therefore a thermal expansion causes a gap reduction \cite{francisco-lopezJPCL2019}. For instance, the pressure coefficient for \ce{GaAs} was experimentally estimated and it yields a positive value of $\sim 80-130$ meV/GPa \cite{chang1984pressure}.

The pressure coefficient can be used to quantitatively estimate the temperature dependence of the band gap of semiconductors, such as black phosphorus \cite{villegas2016NanoLett} and lead halide perovskites \cite{francisco-lopezJPCL2019}. Thus, for penta-\ce{NiN2}, we plot the band gap as a function of the crystal pressure, as shown in Fig.~\ref{fig:eg-pressure}. As a result, we compute a negative pressure coefficient of $-49$ meV/GPa, indicating that this pentagonal 2D semiconductor exhibits an anomalous temperature dependence, which means that its bandgap tends to increase with increasing temperature.



\begin{figure*}[t!]
    \centering
    \includegraphics[width=\textwidth]{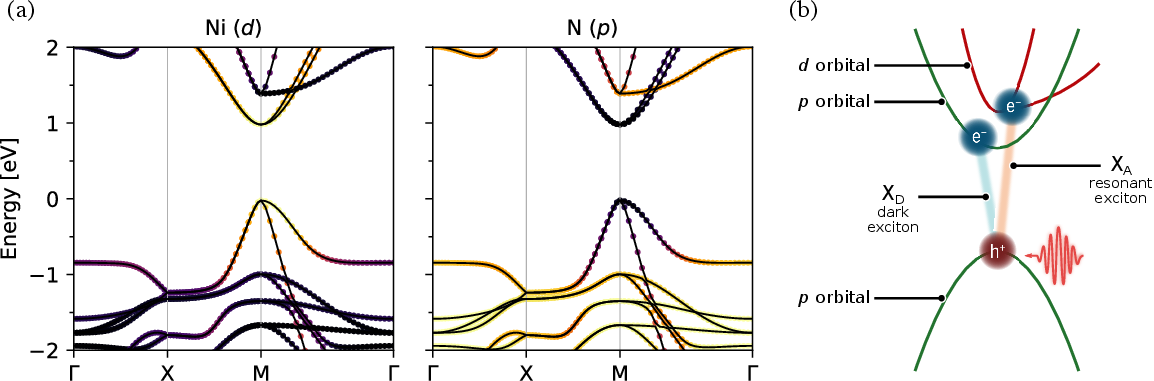}
    \caption{(Color online) Orbital-resolved bandstructures of penta-\ce{NiN2} projected into Ni(\textit{d}) and N(\textit{p}) orbitals. The color scale indicates the magnitude of the projection. The Kohn-Sham eigenvalues were shifted by a scissor operator of 0.97 eV, corresponding to the computed quasiparticle correction within the eigenvalue self-consistent $GW$ (ev$GW$). (b) The scheme of the main optical transitions is depicted, with a representation of the dark (X\textsubscript{D}) and the most intense resonant exciton (X\textsubscript{A}) at the M valley in reciprocal space.}
    \label{fig:fatbands}
\end{figure*}

The optical spectrum of the monolayer penta-{NiN2} including excitonic effects is shown in Fig.~\ref{fig:absorbance-oscillator_pNiN2}\textcolor{blue}{(a)}. Those calculations were carried out by solving the BSE with $evGW$ quasiparticle corrections for the band gap.  The calculated absorption spectra were also obtained at the independent particle (IP) level, for light-polarized in [100] and [010] directions. The most prominent feature is the isotropic optical absorbance yielded from the centrosymmetric structure of penta-\ce{NiN2} at both BSE and IP level of theory. As expected, when the electron-hole interaction is switched on, it induces a red shift and a redistribution of the oscillator strengths. 

An optically forbidden (allowed) dark (bright) excitonic transition can be described according to the very small (high) oscillator strength of optical transitions. The normalized oscillator strengths for the monolayer penta-{NiN2} are shown in Fig.~\ref{fig:absorbance-oscillator_pNiN2}\textcolor{blue}{(b)}. The ground-state exciton X\textsubscript{D} is dark, with a binding energy of 0.34 eV. The most prominent excitonic peak is located at $\sim1.1$ eV, slightly above the quasiparticle band gap within the ev$GW$ approach. Therefore, the exciton X\textsubscript{A} exciton can be characterized as a resonant exciton, which occurs due to the presence of strong many-body interactions \cite{tang2019ResonantExcitonsWSe2}. 

The presence of resonant excitons has been theoretically predicted for graphene \cite{PRL-Louie-resonantExciton-graph}, and it has already been experimentally observed in both graphene \cite{PRL-exp-resonEx-graph} and \ce{WSe2} monolayer \cite{tang2019ResonantExcitonsWSe2}. For the latter, the resonant excitons were identified by combining temperature-dependent high-resolution spectroscopic ellipsometry at low temperatures with high-energy photoluminescence spectroscopy. In contrast to Wannier-Mott excitons, the free-carrier nature of resonant excitons offers potential advantages for optoelectronics, once the separation of the photoexcited carriers is relevant \cite{palummo2023study}.

To gain further insight into the optical spectra, we plot the square modulus of the wave function for the ground-state dark exciton X\textsubscript{D} and the most prominent bright exciton X\textsubscript{A}, as shown in Figs.~\ref{fig:absorbance-oscillator_pNiN2}\textcolor{blue}{(c)} and \ref{fig:absorbance-oscillator_pNiN2}\textcolor{blue}{(d)}, respectively. To obtain these isosurfaces, we fix the hole at a N site and then project the exciton probability amplitude onto the $x{-}y$ plane. We notice that the main and secondary diagonals play important roles in the real-space electron distribution, due to the position of the atoms and the directions of the nitrogen-nitrogen bonds. Regarding diagonal directions, we can verify exciton resonant states for both X\textsubscript{D} and X\textsubscript{A}, so that they have characteristic residual amplitudes that are not significantly decaying away from the hole \cite{liang2012many}. All the main optical transitions occur in the vicinity of the $M$ point, and the exciton X\textsubscript{A} is due to transitions from the valence band maximum (VBM) to the third conduction band (CBM+2).

The selection rules involved in the optical transitions for monolayer penta-{NiN2} were investigated by means of the orbital-resolved band structure. We apply a scissor operator of $0.97$ eV in the PBE band structure, which corresponds to the calculated ev$GW$ quasiparticle gap renormalization. The results for Ni(\emph{d}) and N(\emph{p}) orbitals are shown in Fig.~\ref{fig:fatbands}\textcolor{blue}{(a)}. 

As we can verify, the VBM and CBM originated from the \emph{p} orbitals of N atoms, whereas the CBM+2 is composed of \emph{d} orbitals from Ni atoms. Since the most prominent bright exciton X\textsubscript{A} occurs due to optical transitions from VBM to CBM+2 (see Fig.~\ref{fig:absorbance-oscillator_pNiN2}\textcolor{blue}{(d)}, bottom panel), we can state that the Laporte or orbital selection rule governs the optical transitions in single-layer penta-\ce{NiN2}. Taking into account that its structure is centrosymmetric, the Laporte selection rule establishes that electronic transitions that conserve the wavefunction parity are forbidden, implying that \emph{p} to \emph{p}, or \emph{d} to \emph{d}, transitions should not be observed. The \emph{p} to \emph{d} optical transition verified for exciton X\textsubscript{A} is orbitally allowed with a change in the azimuthal quantum number $\Delta \ell = +1$. On the other hand, transitions from VBM to CBM tend to be forbidden since $\Delta \ell = 0$. The dark and bright excitons in single-layer penta-\ce{NiN2} considering the optical transitions either allowed or forbidden by Laporte rule are summarized in Fig.~\ref{fig:fatbands}\textcolor{blue}{(b)}.


\section{\label{sec:conclusions}Conclusions}
In conclusion, we investigated the optoelectronic
properties of monolayer penta-{NiN2}, a recently synthesized pentagonal 2D semiconductor with promising performance in optoelectronic devices. Employing state-of-the-art parameter-free excited-state calculations, within the $GW$ approach, and by solving the Bethe-Salpeter equation, we have computed the quasiparticle band structure and the optical absorption spectrum including excitonic effects. Employing the eigenvalue self-consistent $GW$ approach, its quasiparticle band gap yielded ${\sim}1$ eV. The ultra-high hole mobility of ${\sim}80\times10^4$ cm$^2$/V$\cdot$s and the negative pressure coefficient for this pentagonal 2D semiconductor contribute to a set of promising potential for high-speed electronic devices. The optical absorption spectrum indicates that the ground-state exciton X\textsubscript{D} is dark, while the most intense optical peak X\textsubscript{A} is a resonant exciton with a free-carrier nature. The optical transitions in monolayer penta-{NiN2} can be explained by means of the Laporte selection rule, providing a comprehensive description of the absorption features and excitonic states in this 2D pentagonal semiconductor. Therefore, we shed light on promising optoelectronic properties of single-layer penta-\ce{NiN2}, paving the way for potential applications using this pentagonal 2D semiconductor. 

\section*{\label{sec:acknowledgements}Acknowledgements}
E.M.Jr. acknowledges the financial support from the Brazilian agency FAPESP, Grant No.~20/13172-8 and 2017/02317-2. This research was supported by resources supplied by the Center for Scientific Computing (NCC/GridUNESP) of the UNESP and the Centro Nacional de Processamento de Alto Desempenho em São Paulo (CENAPAD-SP).

\bibliography{references}

\end{document}


\renewcommand{\thetable}{S\arabic{table}}
\renewcommand{\thefigure}{S\arabic{figure}}
\begin{center}
    { \large\bf\fontfamily{lmss}\selectfont Supplemental Material:}
\end{center}
\title{Many-body effects on the quasiparticle
band structure and optical response of single-layer penta-\ce{NiN2}}
\author{Enesio Marinho Jr.}
 \email{enesio.marinho@unesp.br}
\affiliation{Departamento de Física e Química, Universidade Estadual Paulista (UNESP),\\ Av.\ Brasil, 56, Ilha Solteira, 15385-007 São Paulo , Brazil}
\affiliation{
 Instituto de Física Teórica, Universidade Estadual Paulista (UNESP),\\
R.~Dr.~Bento Teobaldo Ferraz, 271, 01140-070 São Paulo, São Paulo, Brazil.}
\author{Cesar E. P. Villegas}
\email{cesarperezvillegas@gmail.com}
\affiliation{
 Departamento de Ciencias, Universidad Privada del Norte, Lima 15434, Peru}
 \affiliation{
  Instituto de Física Teórica, Universidade Estadual Paulista (UNESP),\\
R.~Dr.~Bento Teobaldo Ferraz, 271, 01140-070 São Paulo, São Paulo, Brazil.}
\author{Pedro Venezuela}
\email{pedrovenezuela@id.uff.br}
\affiliation{Instituto de Física, Universidade Federal Fluminense (UFF),\\ Av.~Gal.~Milton Tavares de Souza, s/n, 24210-346 Niterói, Rio de Janeiro, Brazil.}
\affiliation{
 Instituto de Física Teórica, Universidade Estadual Paulista (UNESP),\\
R.~Dr.~Bento Teobaldo Ferraz, 271, 01140-070 São Paulo, São Paulo, Brazil.}
\author{Alexandre R. Rocha}
\email{alexandre.reily@unesp.br}
\affiliation{
  Instituto de Física Teórica, Universidade Estadual Paulista (UNESP),\\
R.~Dr.~Bento Teobaldo Ferraz, 271, 01140-070 São Paulo, São Paulo, Brazil.}
 

\maketitle

\clearpage
\renewcommand{\thesection}{S-\Roman{section}}
\section{Convergence tests for $GW$ calculations}

\begin{figure}[!ht]
    \centering
    \includegraphics[width=\textwidth]{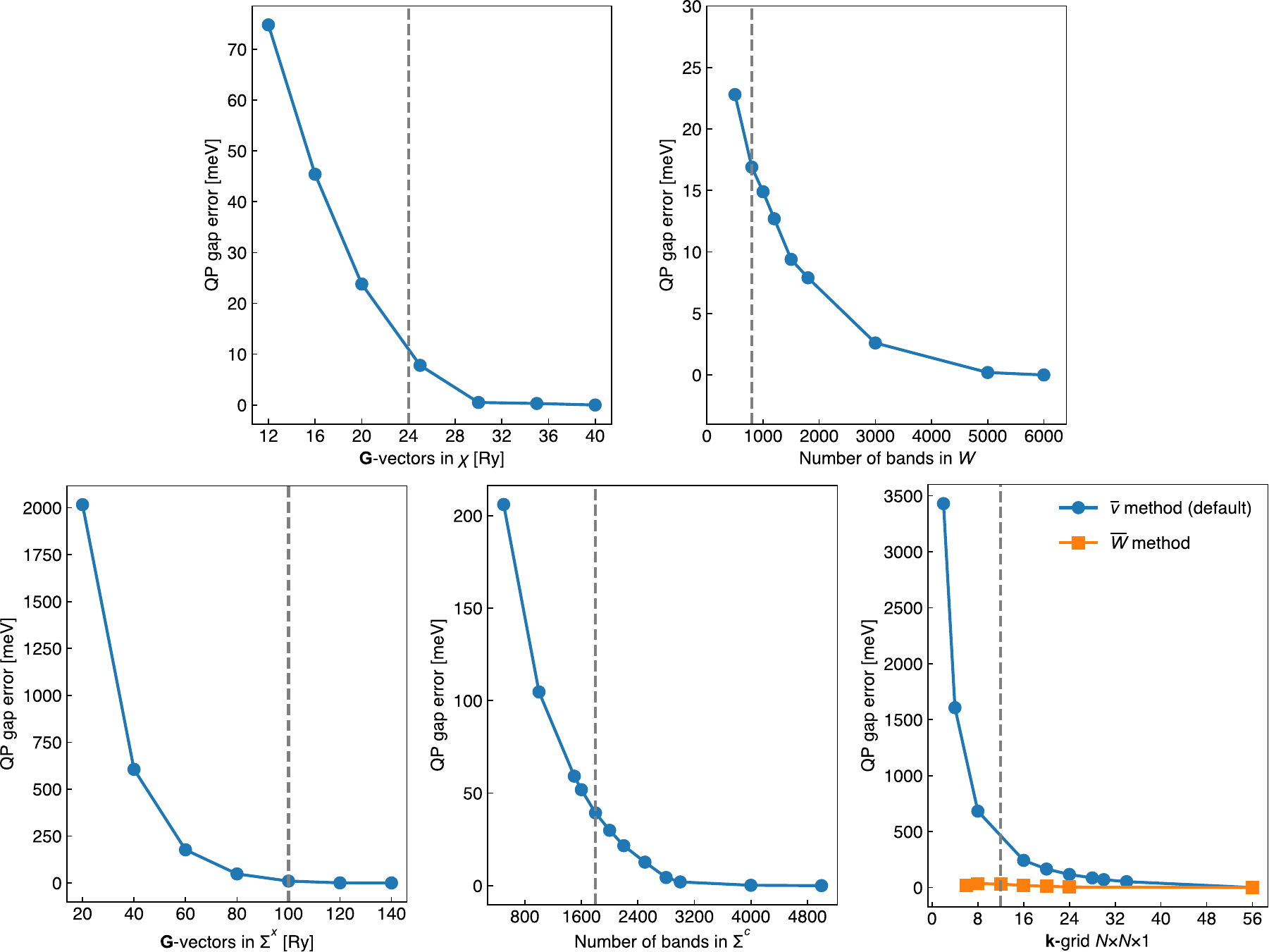}
        \caption{Convergence of quasiparticle correction for the $GW$ band gap of penta-\ce{NiN2} at $M$ $\mathbf{k}$-point as a function of the following features: $\vb{G}$-vectors in the response function (screened interaction cutoff), number of bands in screening $W$, $\vb{G}$-vectors in the exchange part of self-energy $\Sigma^x$, number of bands in correlation part of self-energy $\Sigma^c$, and $\vb{k}$-mesh ($N{\times}N{\times}1$). The QP gap error is estimated by comparing the $G_0W_0$ gap with the one obtained with the largest value for the parameter under consideration.}
\end{figure}

\clearpage
\section{Convergence tests for BSE optical spectrum} 
\begin{figure}[!ht]
    \centering
    \includegraphics[width=\textwidth]{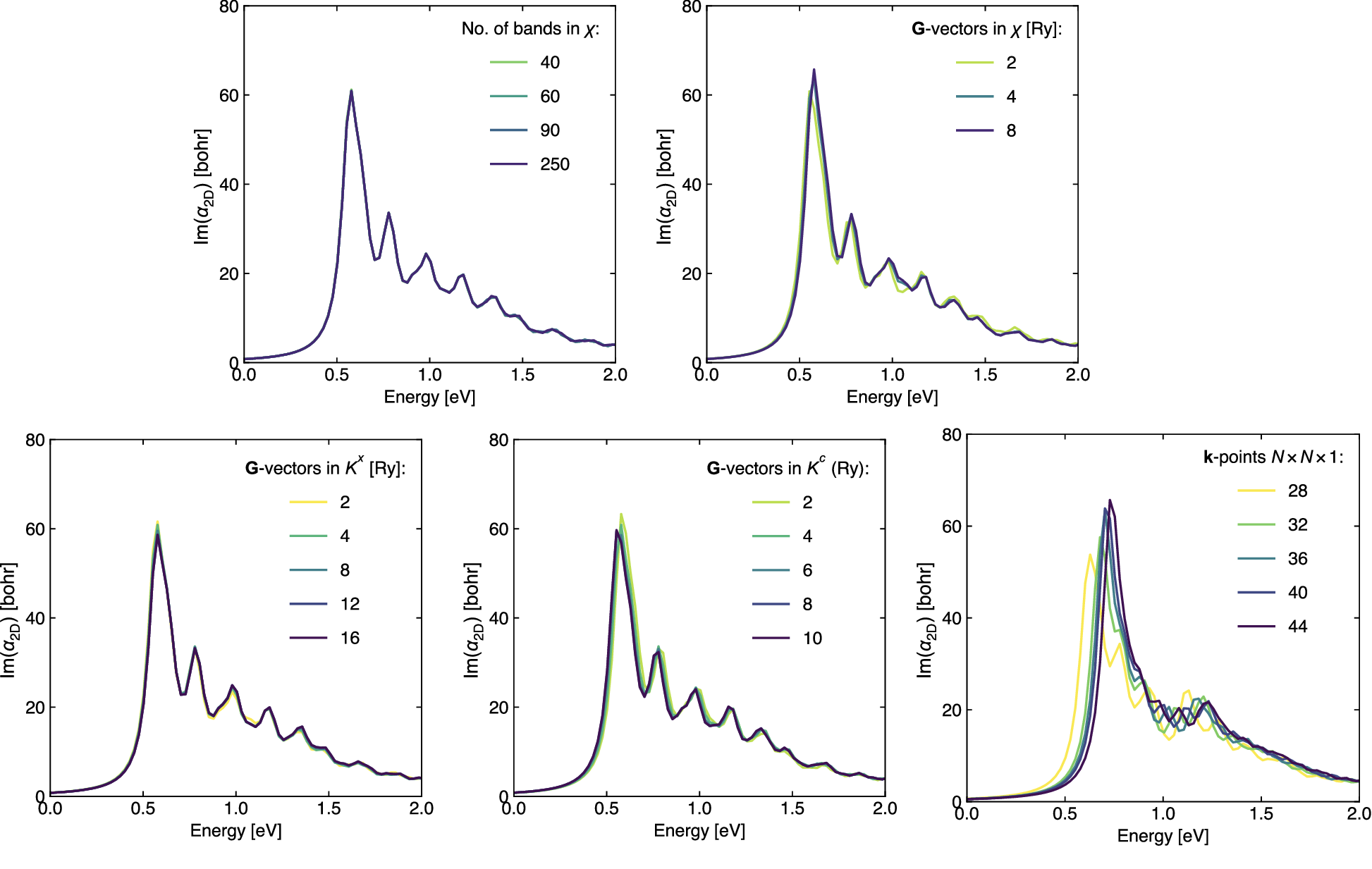}
        \caption{Convergence of the BSE imaginary part of the polarizability for penta-\ce{NiN2} as a function of: number of bands and $\vb{G}$-vectors in response function (in screening $W$), $\vb{G}$-vectors in exchange and in correlation part of BSE kernel ($K^x$ and $K^c$), and $\vb{k}$-grid sampling ($N{\times}N{\times}1$).}
\end{figure}
